\documentclass[
    a4paper,
    prb,
    bibnotes,
    twocolumn,
    showpacs,
    superscriptaddress,
    groupedaddress
]{revtex4}

\usepackage{array}
\usepackage{graphicx}
\usepackage{amssymb}
\usepackage{amsmath}
\usepackage{fullpage}
\usepackage{hyperref}
\usepackage{paralist}
\usepackage{subfigure}
\usepackage{mathdots}
\usepackage{float}
\hypersetup{
    colorlinks   = true,
     citecolor    = blue,
     linkcolor    = blue,
     citecolor    = blue,
     filecolor     = blue,
     urlcolor	      =	blue  
}

\setcitestyle{square}
\newcommand{\bs}{\boldsymbol}
\newcommand{\beq}{\begin{equation}}
\newcommand{\eeq}{\end{equation}}
\newcommand{\bra}[1]{\langle #1 |}
\newcommand{\ket}[1]{| #1 \rangle}

\newcommand{\beqa}{\begin{eqnarray}}
\newcommand{\eeqa}{\end{eqnarray}}

\newcommand{\etal}{\mbox{\textit{et al.}}}

\newcommand{\al}[1]{\begin{align}#1\end{align}}

\newcommand{\Eqref}[1]{Eq.~(\ref{#1})}
\newcommand{\Figref}[1]{Fig.~\ref{#1}}

\begin{document}
\title{Scanning tunneling microscopy current from localized basis orbital density functional theory}
\date{\today}

\author{Alexander \surname{Gustafsson}}
\email{alexander.gustafsson@lnu.se}
\affiliation{Department of Physics, Linnaeus University, 391 82 Kalmar, Sweden}
\author{Magnus \surname{Paulsson}}
\email{magnus.paulsson@lnu.se}
\affiliation{Department of Physics, Linnaeus University, 391 82 Kalmar, Sweden}

\begin{abstract}
We present a method capable of calculating elastic scanning tunneling microscopy (STM) currents from localized atomic orbital density functional theory (DFT). To overcome the poor accuracy of the localized orbital description of the wave functions far away from the atoms, we propagate the wave functions, using the total DFT potential. From the propagated wave functions, the Bardeen's perturbative approach provides the tunneling current. To illustrate the method we investigate carbon monoxide adsorbed on a Cu(111) surface and recover the depression/protrusion observed experimentally with normal/CO-functionalized STM tips. The theory furthermore allows us to discuss the significance of $s$- and $p$-wave tips.
\end{abstract}
\pacs{68.37.Ef, 33.20.Tp, 68.35.Ja, 68.43.Pq}
\maketitle

\section{Introduction}
Scanning tunneling microscopy has made large strides since its conception 35 years ago. Advances include studies of detailed atomic structure including the shapes of molecular orbitals\cite{Gross2011}, atomic manipulation\cite{Eigler1990}, and inelastic spectroscopy from vibrational\cite{Stipe1998} and magnetic excitations\cite{Hirjibehedin2007}. Theoretically, the standard methods of Bardeen\cite{Bardeen1961} and the Tersoff-Hamann\cite{Tersoff1985} approximation have served to model and provide understanding of the STM measurements. The Tersoff-Hamann approach to model STM experiments from first principles calculations has provided a clear understanding of many phenomena\cite{Persson2002}. However, with the emergence of non $s$-wave functionalized STM tips\cite{Gross2011}, e.g., CO functionalized tip, refined methods are required such as the Bardeen method\cite{Paz2006,Rossen2013,Bocquet1996,Teobaldi2007,Zhang2014} or Chen's derivative method\cite{Chen1990, Mandi2015, Mandi2015b} with a proper description of the tip states. 

Advances in the related field of conduction through nanoscale devices has made great progress using non-equilibrium Green's function methods\cite{Paulsson2010} 
However, these methods are in practice not directly applicable to STM modelling since (i)
scanning the STM tip over the surface results in many costly computations, and (ii) the prevalence of localized basis set DFT calculations to describe the electronic structure, which cannot accurately capture the tunneling at large distances. These difficulties to use large scale DFT calculations  with localized basis set to model STM images has given rise to several methods to improve the modeling capabilities\cite{Rossen2013,Paz2006,Zhang2014}. 

In this work we pursue a similar methodology as Paz \etal\cite{Paz2006} where the electronic states of the tip and sample sides 
are calculated accurately close to the tip and sample sides using the localized basis DFT method \textsc{Siesta}\cite{Soler2002}. 
These states are then propagated in the vacuum region to provide accurate states 
to use in the Bardeen formalism. However, instead of assuming a flat potential in the vacuum region we utilize the total DFT potential landscape to propagate the wave functions into the vacuum region, and thereby, by the Bardeen method, construct first principles STM images. To benchmark our method, we focus on describing the current dip over a carbon monoxide molecule (CO) adsorbed on the Cu(111) surface\cite{Heinrich2002}, which has been well studied experimentally as well as being used to provide larger resolution when the STM tip is functionalized by the CO molecule\cite{Bartels1997}.\\

\section{Theory}
\begin{figure}[t!]
	\includegraphics[width=\columnwidth]{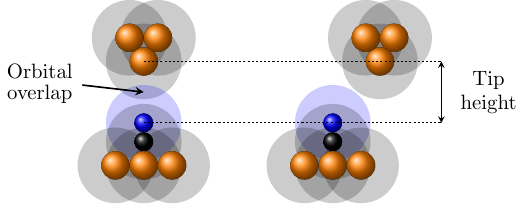}
	\caption{Illustration of CO adsorbed on a Cu surface and Cu tip. The finite range basis orbitals used in \textsc{Siesta}, shaded region, underline the difficulty of describing the  tunneling current with a localized basis.\label{figOverlap}}
\end{figure}

Our theoretical framework relies on the well known Bardeen\cite{Bardeen1961} approximation. In the small current limit, time-dependent perturbation theory using a Fermi's golden rule like formalism, alternatively non-equilibrium Green's function theory\cite{Pendry1991,Corbel1999}, gives the tunneling current as:
\beq
	I=\frac{2\pi e}{\hbar}\sum_{t,s}\left[f(\varepsilon_t)-f(\varepsilon_s)\right]\left|M_{ts}\right|^2\delta(\varepsilon_t-\varepsilon_s+eV)\label{STMcurrent},
\eeq
where $f(\varepsilon_{t,s})$ are the Fermi-Dirac functions, $\varepsilon_{t,s}$ the energy levels of tip and substrate relative respective chemical potential, and $V$  the applied bias. The matrix element, $M_{ts}$, couples a tip state, $\varphi_t$, to a substrate state, $\varphi_s$, by the expression
\beq
	M_{ts}=-\frac{\hbar^2}{2m}\int_S\textrm{d}S\cdot \big[\varphi_t^*(\bold{r})\nabla\varphi_s(\bold{r})-\varphi_s(\bold{r})\nabla\varphi_t^*(\bold{r})\big]\label{Mts},
\eeq
where the integral is evaluated on any surface  in between the adsorbate and tip. In this article, we focus on the low bias conductance, where only the tip and substrate wave functions close to the Fermi energy are needed. Henceforth, when referring to the tunneling current in the small bias limit, we simply use $I=GV$, where $G=G_0T$, and $T$ is the transmission probability, $T=4\pi^2|M_{ts}|^2$, for a specific tip-substrate combination, and $G_0=e^2/(\pi\hbar)$ is the conductance quantum. Generalizations to investigate bias dependent tunneling should be straightforward. Since the integration surface is far from either the tip or substrate, the low electron density in addition to the use of finite range basis orbitals in \textsc{Siesta}, c.f., \Figref{figOverlap},  the  wave functions are poorly described in the vacuum region. In contrast, the wave functions close to the atomic nuclei can be accurately and efficiently calculated by \textsc{Siesta}. STM images can thus be calculated by propagating these accurate wave functions (close to the nuclei) into the vacuum region. 

\subsection{Propagation of wave functions}
\begin{figure}[t!]
	\includegraphics[width=\columnwidth]{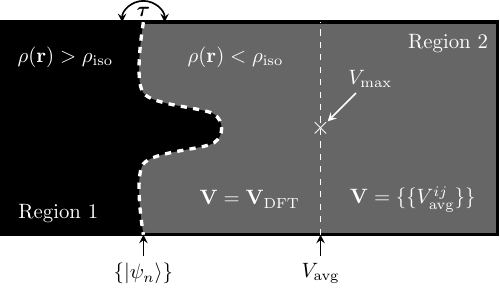}	
	\caption{2D illustration of the device (full region), substrate (black) and vacuum (gray) regions. The dashed white line illustrates the constant density surface, $\rho_{\textrm{iso}}$ from which the substrate states, $\{\ket{\psi_n}\}$, are propagated into the vacuum region.\label{fig2Dcartoon}}
\end{figure}

The \textsc{Siesta} method solves the Kohn-Sham equations using norm conserving pseudopotentials and a localized basis set. Outside the range of the pseudopotentials, the Kohn-Sham orbitals obey, at a specified energy, a Schr\"odinger like equation with a local potential, which makes it conceptually straightforward to propagate a known wave function from a surface into the vacuum region. The substrate and tip sides are treated in the same way, and to simplify the presentation we henceforth only describe the details starting from substrate side. To propagate the wave functions in real space, we start from a charge density iso-surface, see \Figref{fig2Dcartoon}, chosen close to the substrate but outside of the radii of the pseudopotentials. In the vacuum region, the Kohn-Sham equations (Rydberg units) therefore read $(-\nabla^2+V_{\textrm{tot}}) \Psi =\varepsilon_{\textrm{F}} \Psi$ where the total potential ($V_{\textrm{tot}}$) contains the Coulomb, Hartree, and exchange-correlation terms. For efficiency, the calculation cell actually contains both substrate and tip situated far enough away from each other to give a negligible current. To remove the tip potential we therefore find the real space grid slice containing the maximum value of the total DFT potential value and set the total potential further away from the substrate to the average value of $V_{\textrm{tot}}$ at that height above the substrate. In order to use this modified DFT potential, a sufficiently large vacuum gap is required in order to capture the work function properly. 

The substrate and tip states are obtained for the device region coupled to semi-infinite electrodes using \textsc{Transiesta}\cite{Brandbyge2002}, which is a non-equilibrium Green's function method built on the \textsc{Siesta} method. The states at the Fermi energy are found in the \textsc{Siesta} basis set by diagonalizing the partial spectral function\cite{Paulsson2007}, i.e., the spectral function of the substrate (tip) side $A_{s(t)}=G_d\Gamma_{s(t)}G_d^{\dagger}=\sum_n \ket{\widetilde{\psi}_n}{\lambda_n}\bra{\widetilde{\psi}_n}$, where $G_d$ is the retarded Green's function of the device region, and $\Gamma_{s(t)}$ is the broadening of the substrate (tip). Although the device region can be arbitrarily large, only a few (of the order of 10 in our examples) eigenvalues are non-zero and the corresponding eigenvectors give the scattering states when properly normalized\cite{Paulsson2007}, 
\beq
	\ket{\psi_n}=\sqrt{\frac{\lambda_n}{2\pi}}\ket{\widetilde{\psi}_n}\label{eqDiagA},
\eeq
where $\ket{\psi_n}$ are the energy normalized  scattering states. Any unitary transformation of a scattering state, e.g., sign flip or rotation, does not affect the physical observables. Note that the scattering states are neither basis orbitals nor current eigenfunctions. They are formed by incoming Bloch states in the semi-infinite electrodes which are almost totally reflected. The number of states is therefore determined by the number of Bloch states in the leads at the Fermi energy. Owing to the finite transmission probability, the scattering states consist of a real and an imaginary part, where the real part is dominant (by a factor $\sim10^3$) due to almost total reflection of the waves. Hence, in the discussion below, only the real part is considered when visualizing the wave functions. To propagate these states into the vacuum region, the wave functions are evaluated on the same real space grid as used by \textsc{Siesta}. 

The real space wave functions can then be computed using the Green's function formalism. Here we use the finite difference (FD) method to discretise the device region\footnote{A version using the finite element method is under development.}, i.e., the gray and black regions in \Figref{fig2Dcartoon}. The FD Hamiltonian is a sparse tight-binding Hamiltonian which, by separating the full device region into subspaces, can be written as
\beq
\bold{H}=\left[\begin{array}{cc}\bold{H}_1&\bs{\tau}\\ \bs{\tau^{\dagger}}&\bold{H}_2\end{array}\right],
\eeq
where subspace 1 contains the volume with the high charge density (close to the atoms), and subspace 2 is the vacuum region. The matrix $\bs{\tau}$ describes the coupling elements between the two regions.

By means of the Dyson equation, the wave functions inside region 1, $\ket{\psi_n}$ (known from the \textsc{Siesta} calculation), relate to the propagated wave functions, $\ket{\varphi_n}$, in the vacuum region as
\beq
	\ket{\varphi_n}=\bold{G}_0\bs{\tau}\ket{\psi_n}\label{eqMain},
\eeq
where $\bold{G}_0=(\varepsilon_{\textrm{F}}\bold{I}-\bold{H}_2-\bs{\Sigma}_R)^{-1}$ is the isolated Green's function of the vacuum region, optionally connected to a semi-inifinite continuation on the right hand side by the self-energy $\bs{\Sigma}_R$. The Hamiltonian, $\bold{H}_2$, contains the FD Laplacian of the vacuum (region 2, \Figref{fig2Dcartoon}), and the total potential, including the modified vacuum part far away from the substrate.  The total potential and charge density are readily available in real space from the \textsc{Siesta} code, and we therefore only need to specify the boundary conditions to be able to propagate the wave functions. Furthermore, calculation of the self-energy on the vacuum boundary, $\bs{\Sigma}_R$, can safely be omitted if the device region is large, since the propagated modes vanish exponentially fast in the vacuum region (i.e., near total reflection). Inclusion of the vacuum self-energy is therefore optional as long as the device region is large enough, and will be omitted in the following. We compute the modes in the vacuum region, $\ket{\varphi_n}$, by algebraically solving the (sparse) linear system of equations,
\beq
	\left(\varepsilon_{\textrm{F}}\bold{I}-\bold{H}_2\right)\ket{\varphi_n}=\bs{\tau}\ket{\psi_n}\label{eqPmodes},
\eeq 
Note that solving the system of linear equations is much less demanding than performing a matrix inversion to obtain $\bold{G}_0$, allowing for larger systems to be considered. 

Assuming that the calculated wave functions are unchanged by translation of the tip relative to the substrate allows us to efficiently calculate 
the STM current image as a convolution in two dimensions,
\al{
M_{ts}(\bold{R})=-\frac{\hbar^2}{2m}\int_S\textrm{d}S&\cdot \big[\varphi_t^*(\bold{r}+\bold{R})\nabla\varphi_s(\bold{r})\nonumber\\
&-\varphi_s(\bold{r})\nabla\varphi_t^*(\bold{r}+\bold{R})\big]\label{MtsR},
}
where the convolution is performed efficiently using Fast Fourier Transforms (FFT)\cite{Paz2006,Zhang2014}. Shifting the wave functions in the $\hat{z}$-direction is easily accomplished and the computational cost negligible.

\subsection{Computational details}
We use the \textsc{Siesta}\cite{Soler2002} DFT program to obtain the relaxed geometry of the supercell. The substrate consists of five $4\times4$ Cu-layers forming the Cu(111) geometry, with lattice constant 2.57 \AA~hence giving images of dimensions $10.3\times10.3$ \AA. The molecule, the two top layers of the substrate, and the tip atom are relaxed until the residual force is less than 0.04 eV/\AA. The tip consists of three similar layers with a pyramidal tip apex with four Cu-atoms attached to the underside of the tip slab. A part of the relaxed supercell geometry is visualized in \Figref{figCurrent}. The maximal range of the atomic orbitals are 4.8, 3.0 and 3.9 \AA~for Cu, O, and C respectively. Periodic boundary conditions are imposed in all spatial dimensions. The computations were performed with the PBE GGA functional\cite{Perdew1996}, SZP(DZP) basis set for Cu(CO), a 200 Ry mesh cutoff energy for real space grid integration, and $4\times4$ \textbf{k}-points in the surface plane. Non-equilibrium wave functions are calculated by the \textsc{Transiesta}\cite{Brandbyge2002} and \textsc{Inelastica}\cite{frederiksen2007} modules by adding three(six) layer electrodes at the substrate(tip) side of the supercell.  

For the real space wave function propagation we have found that the real space grid given by \textsc{Siesta} (200 Ry cutoff) is unnecessarily fine, and in the wave function propagation the grid coarseness were doubled (corresponding to a 50 Ry cutoff in \textsc{Siesta}). This reflects that the potential and the wave functions in the vacuum region change slowly compared to closer to the atoms. Furthermore, the propagation, and STM image computation time is reduced from $\sim$hours (200 Ry) to $\sim$minutes (50 Ry)\footnote{The largest computational hurdle is to obtain the relaxed DFT solution using \textsc{Siesta} and \textsc{TranSiesta} which requires tens of CPU days for the systems studied here.}.

The grid sizes used below consist of $40\times40$ points in the plane of the substrate and approximately 50 points along the transport direction. The density on the isosurface, $\rho_{\textrm{iso}}$, is chosen so that the this surface lies just outside the radii of the pseudopotentials. Henceforth, we will use $\rho_{\textrm{iso}}=5\times10^{-3}$ Bohr$^{-3}$Ry$^{-1}$, and we have verified that small changes in this parameter do not affect the results. In this study, the STM images were found from the scattering states in the $\Gamma$-point for computational reasons. Further extensions of the code to include \textbf{k}-point sampling is being considered.

\section{Results}
We demonstrate the theory by primarily considering a CO molecule adsorbed on a top site of a clean $4\times4$ atom Cu(111) surface with a pyramidal shaped tip apex, consisting of four copper atoms, see \Figref{figOverlap}. A comparison between direct use of the scattering states computed from the localized \textsc{Siesta} basis, and the modes found by \Eqref{eqPmodes} will first be examined to highlight the advantages of our theory. This reveals the characteristic depression over the CO molecule in the tunneling current, which will be analysed in terms of the individual wave functions, and compared with the STM image obtained with a CO-terminated tip. The tip-height is henceforth defined as the core-core height difference (along $\hat{z}$) between the outermost tip apex atom and its closest atom adsorbed on the substrate, see \Figref{figOverlap}.

\subsection{Comparison between localized and propagated wave functions}
\begin{figure}[htb!]
	\includegraphics[width=\columnwidth]{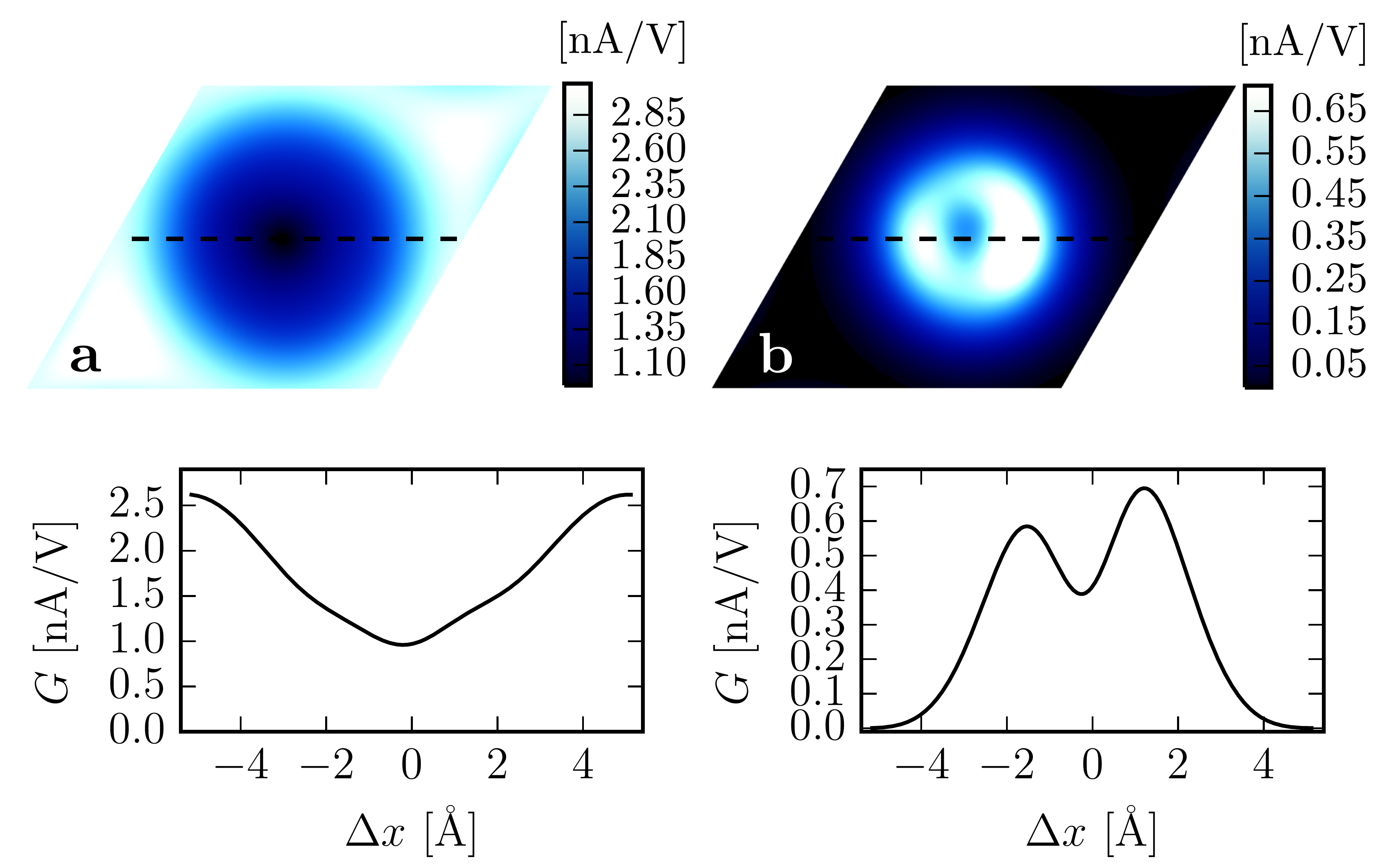}
	\caption{Comparison of the calculated constant height STM image at 5.0 \AA, computed by means of a) the propagated, and b) \textsc{Siesta} wave functions. Bottom panel shows the cross sections along the dashed lines. \label{Fig0}}
\end{figure}

\begin{figure*}[htb!]
	\includegraphics[width=\textwidth]{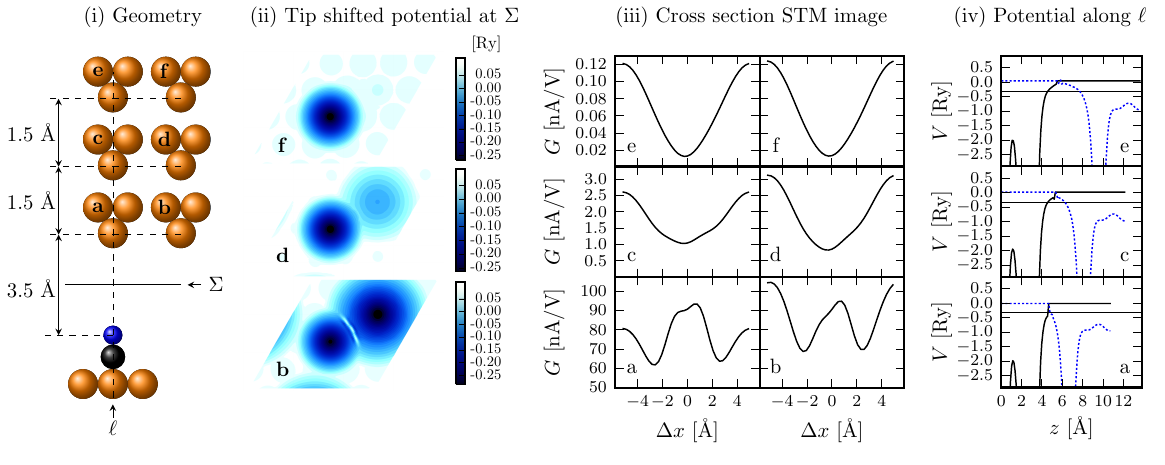}
	\caption{STM image dependence on the substrate-tip distance and lateral displacement of the tip. (i) Illustrations of the six geometries, each computed by a separate DFT calculation. (ii) the total potential at the separation surface, $\Sigma$, for the 3.5 \AA~system, in (i). (iii) Constant height STM image cross sections for the six DFT calculations. (iv) total potential energy along the bonding axis of the CO molecule where the black (dashed blue) line is the potential used in propagation of the wave functions from the substrate (tip) side, and the horizontal black line is the Fermi energy. \label{TipShift}}
\end{figure*}

To illustrate the limitations of using the localized orbitals to describe the wave functions in the vacuum gap, we use \Eqref{STMcurrent} for both the wave functions found directly from \textsc{Siesta}, and compare with using the propagated modes. \Figref{Fig0} displays the constant tip-height (5.0 \AA) tunneling current showing the expected dip in current over the CO molecule using the propagated wave functions (\Figref{Fig0} a), and the failure of the \textsc{Siesta} wave functions to reproduce the dip (\Figref{Fig0} b).

Since the \textsc{Siesta} basis orbitals are exactly zero outside a certain range, the scattering states calculated in the \textsc{Siesta} basis have a vanishingly small overlap when the tip is moved away laterally from the molecule. Over the molecule this overlap is increased, and the CO molecule therefore appears as a protrusion in the tunneling current contrary to the characteristic dip seen experimentally. This can be remedied, at least partially, by using longer range basis functions\cite{Zhang2014}. Here, we instead propagate the \textsc{Siesta} wave functions into the vacuum region and obtain a more accurate representation of the scattering states, especially away (laterally) from the molecule. This provides an overall increase in the tunneling current. Additionally, as will be discussed in more detail below, the characteristic dip over the CO molecule is recovered.

\subsection{Substrate-tip distance}
The basic assumption in the Bardeen approximation is that the potential of the tip does not perturb the wave functions from the substrate and vice versa\cite{Bardeen1961}. To investigate this assumption we carried out calculations at three different substrate-tip distances with the tip above the CO molecule and with the tip laterally displaced from the molecule, see \Figref{TipShift} (i).

Visualizing the DFT total potential at a constant height above the substrate, see \Figref{TipShift} (ii), clearly shows the effect of the tip on the potential at tip-heights below 5 \AA, e.g., \Figref{TipShift} (ii) (f) shows the potential of the molecule and underlying Cu lattice while (d) and (b) clearly show the perturbation of the potential from the tip. The perturbation of the tip potential on the substrate wave functions severely modify the simulated STM images for distances below 5 \AA, see \Figref{TipShift} (iii), and gives differences between the calculated STM images even when the tip is shifted laterally at the same height. Furthermore, the approximation to use a flat potential outside the maximum potential, see \Figref{TipShift} (iv) is also questionable for tip-heights below 5 \AA~since the height of the potential will give the exponential decay into the vacuum region. 

In summary, these calculations demonstrate the need for a relatively large vacuum gap in order to minimize the impact from the tip potential when calculating 
the substrate wave functions and vice versa. In the following calculations we keep the tip-heights larger than 5 \AA. However, this might, in some cases, be an artificially large distance when comparing with experimental STM images. Other possible solutions might be to introduce the tip potential as a perturbation, although this would necessitate the recalculation of the perturbed wave functions at each lateral displacement of the tip.

\subsection{Wave function analysis}
\begin{figure*}[htb!]
	\includegraphics[width=\textwidth]{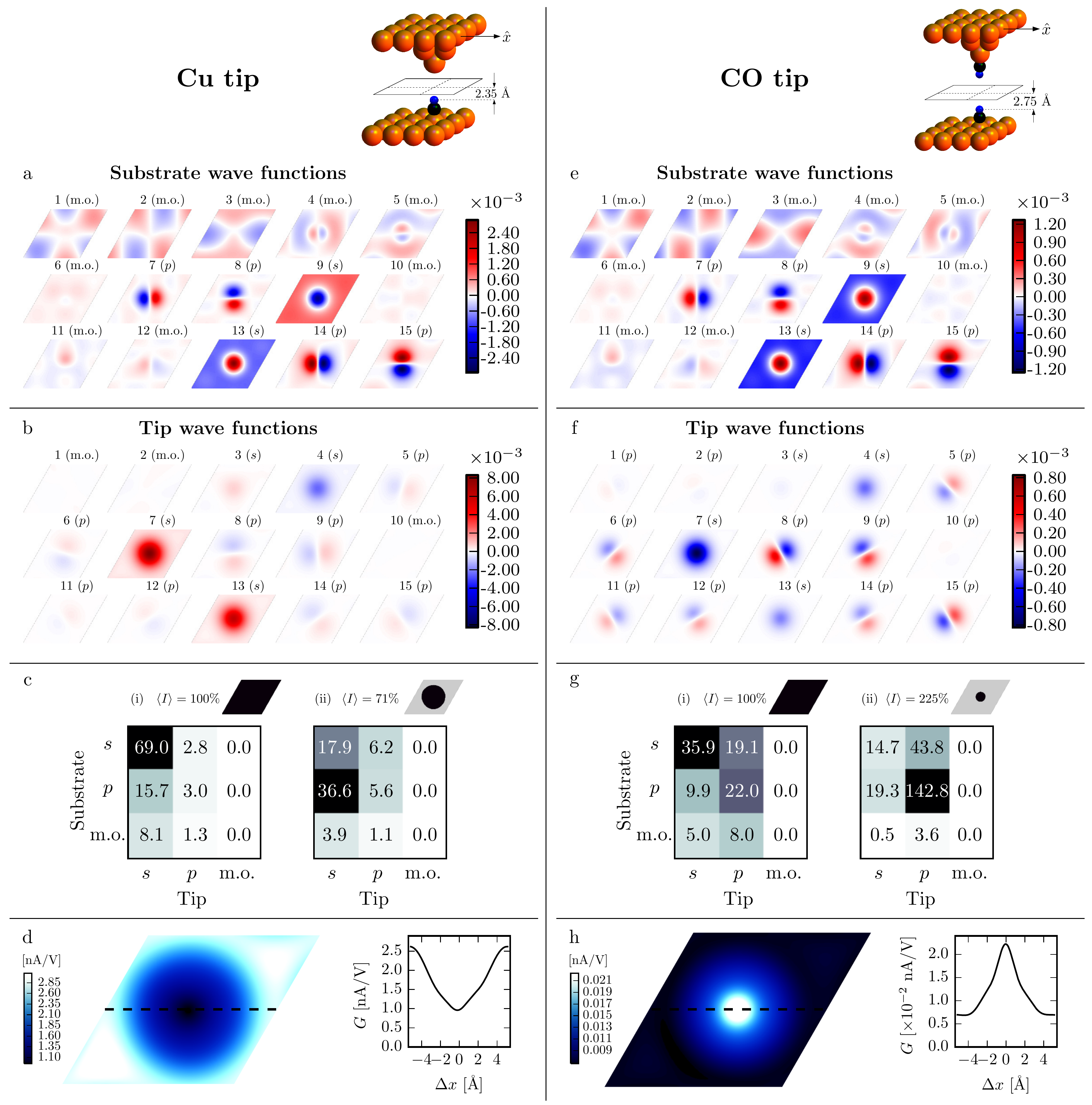}
	\caption{Left column (a, b, c, and d) shows the substrate and tip wave functions for the pyramidal copper ($s$-wave) tip, evaluated at the separation surface in the vacuum region (5.0 \AA~tip-height). The relaxed geometry figure displays the vertical distance between the oxygen atom and the separation plane, and the scan direction for the cross section figures throughout. The matrix in c shows the average tunneling current for the $s$-, $p$-, and miscellaneous wave (m.o.) combinations. The average is taken over tip positions over (i) the full cell and (ii) the area over the molecule, where the details are given in the text. Figure (d) shows the resulting constant height STM image and its cross section along the dashed line. The right column (e, f, g, and h) shows the results with a CO functionalized ($p$-wave) tip at 5.2 \AA~tip-height. The scalebar units in a, b, e, and f are (Bohr$^3$Ry)$^{-1/2}$.\label{figCurrent}}
\end{figure*}

\begin{figure}[t!]
	\includegraphics[width=\columnwidth]{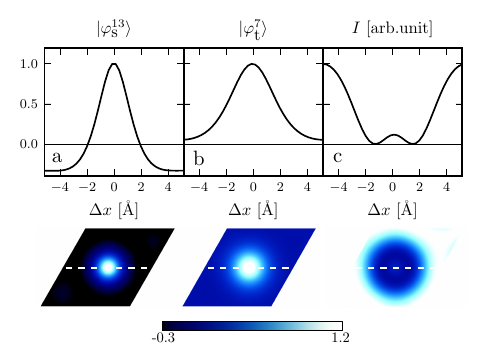}
	\caption{Panel a and b show the most conducting substrate and tip wave function combination in a cross section profile over the CO molecule. Panel c displays a cross section profile of the tunneling current originating from these two states. Below each cross section its 2D image is shown, where the scanning direction is indicated by a dashed line, i.e., along an atom row of the Cu(111) surface, see \Figref{figCurrent}. Peak heights are normalized.\label{figInterference}}
\end{figure}

To gain further insights of the tunneling current contributions to the constant height STM image, it is instructive to look at the various propagated wave functions at the surface on which the integral, $M_{ts}$, c.f., \Eqref{Mts}, is evaluated, see \Figref{figCurrent}. The gradients in the integral can, to zeroth order, be approximated by an exponential decay in the $\hat{z}$-direction, which means that the integral $M_{ts}$ is proportional to the overlap of the wave functions on the integration surface. To separate the contributions from the different types of channels we further separate the contributions of $s$- and $p$-like scattering states from the substrate and tip. They are referred to as $s$- and $p$-waves henceforth, not to be confused with $s$- and $p$-orbitals of individual isolated atoms. The classification into $s$- and $p$-waves is simply done by visual inspection of the integration surface (see \Figref{figCurrent} a, b, e, and f). Note that this means that $p_z$-waves are classified as $s$-waves as they are radially symmetric in the $x$-$y$ plane. Scattering states which are not clearly $s$- or $p$-waves in the $x$-$y$ plane are denoted miscellaneous orbitals (m.o.). This pragmatic classification is sufficient for the current system. However, any unitary mix of the propagated wave functions leaves the physical observables unchanged, and should be considered if the important scattering states have no clear symmetry. This means that, e.g., a sign flip of a particular scattering state does not affect the results. The contributions from the scattering states vary with the tip-position, and we therefore present the average with respect to tip position over the full image or over the molecule. That is, $\left\langle\sum_{s,t\in\left\{s,p,mo\right\}}\left| M_{ts} \right|^2\right\rangle_A$, where the area $A$ is either the full image, \Figref{figCurrent} c (i) and g (i), or the central region over the molecule delimited by the average of maximum and minimum currents, \Figref{figCurrent} c (ii) and g (ii). The resulting plot of the the $\left| M_{ts} \right|^2$ matrix, \Figref{figCurrent} c and g, separates the contributions from the $s$- and $p$-wave scattering states and allows us to discuss the origin of the STM contrast.

\subsubsection{Cu-terminated tip}
For the Cu-terminated tip, the main tunneling current (69\%) originates from a few $s$-waves from both substrate (mainly states 9 and 13) and tip side (7 and 13), see left part of \Figref{figCurrent} c), whereas, the $p$-waves play a minor role. We further note that the tip and substrate scattering states contributing to the current (substrate states 9 and 13, and tip states 7 and 13) are virtually identical apart from a constant. It seems likely that a unitary transformation of the wave functions can provide an even simpler picture, where the main current is carried by a single state substrate and tip state, c.f., current eigenchannels\cite{Paulsson2007}.

\begin{figure}[htb!]
	\includegraphics[width=\columnwidth]{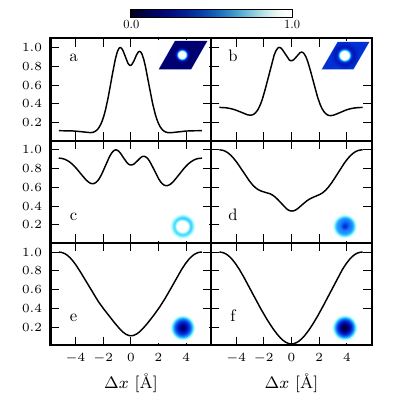}
	\caption{Cross section plots of the normalized local density of states from the substrate, CO@Cu(111), at distance a. 2.5, b. 3.4, c. 4.3, d. 5.2, e. 6.1, and f. 7.0 \AA~above the oxygen molecule. The insets show the 2D images.\label{figLDOS}}
\end{figure}

The average of the current and current matrix over the central area close to the molecule show a clear decrease in the average current, see \Figref{figCurrent} c (ii). Since the Cu tip gives a lower current over the molecule, the average over tip positions close to the molecule is lower (71\%) compared to the average over the full cell, \Figref{figCurrent} c (i) (100\%). The main part of the current is still carried through the $s$-wave tip states although the current going through the substrate $s$-wave decrease substantially, and the importance of the $p$-waves increase markedly. As neither the amplitude, nor the sign of the wave functions are clearly visible in \Figref{figCurrent}, a cross section of the most transmitting mode combination is shown in \Figref{figInterference}. This combination, i.e., $\{\ket{\varphi_\textrm{s}^{13}},\ket{\varphi_\textrm{t}^{7}}\}$ (giving 24\% of the total tunneling current), reveals a sign change for the substrate mode, whereas the sign is strictly positive for the tip mode. The sign change implies that the overlap of the wave functions decreases when the tip is centered over the molecule, and gives, as an interference effect, the dip in tunneling current, see \Figref{figInterference}. Note that the overlap of the two wave functions in  \Figref{figInterference} a and b seems to have a maximum over the molecule if the overlap is calculated in 1D. However, in 2D the resulting overlap is shown in  \Figref{figInterference} c, where the overlap is lower when the tip is centered over the molecule. The difference between the 1D and 2D overlap can be understood as the increase of the area element for increasing $r$ in a polar coordinate system. The depression in the STM image arises simply by interference from the change of the sign of the dominating substrate scattering state. We further note that the scattering state amplitudes are non-zero laterally away from the molecule providing a larger current when the tip is shifted away from the molecule.  

For completeness, the substrate LDOS is shown in \Figref{figLDOS} for several vertical distances. In the Tersoff-Hamann approximation\cite{Tersoff1985}, or the more general Chen's derivative rules\cite{Chen1990}, the current is simply given by the local density of states (LDOS) at the tip position for an $s$-wave tip. At tip-heights $> 5$ \AA~(\Figref{figLDOS} d, e, and f) the LDOS resemble the calculated STM image, c.f., \Figref{figCurrent}. However, in the Bardeen approximation, the integration surface is approximately in the middle of the gap, see \Figref{figCurrent}, where the LDOS does not resemble the STM images (\Figref{figLDOS} a,b, and c). In addition, imaging with a non s-wave tip, e.g., CO-terminated tip, makes the interpretation more difficult.

\subsubsection{CO-terminated tip}
Using a CO-terminated tip increases the importance of the tip $p$-waves and can increase the STM resolution\cite{Gross2011,Garcia2011,Paulsson2008}. Previous studies have introduced $p$-wave tips by expanding the tip-orbitals using Chen's derivative method\cite{Mandi2015} or setting the mix between tip $s$- and $p$-waves by hand\cite{Gross2011}. In addition, the importance of the $p$-waves for the CO-functionalized tip has been seen theoretically in calculations of inelastic electron tunneling spectra\cite{Garcia2011,Paulsson2008}. In our case, the current matrix analysis allows us to analyse the contributions directly from the DFT-calculations of the tip and substrate states. 

For the CO-terminated tip our calculations show the reversal of the STM contrast with a conductance peak  above the CO molecule, c.f., \Figref{figCurrent} h, in agreement with experiments\cite{Bartels1997}. The propagated wave functions and the current contributions from the $s$- and $p$-waves are shown in the right column in \Figref{figCurrent} g. Due to the contrast inversion (compared to the Cu tip, \Figref{figCurrent} d), the average current inside half the maximum of the final STM image, \Figref{figCurrent} g (ii), is higher (225\%) compared to when scanning the full cell (100\%). As with the Cu tip, the various combinations of $s$-waves yield a current dip. However, the relative intensity for these $s$-waves are significantly weaker for the CO tip, whereas the $p$-wave intensities play a dominant role, especially over the molecule, as seen from the current matrix.  Away from the molecule, the $p$-channels are not as dominant but still significant in the resulting STM current. For the $p$-wave tip states, see \Figref{figCurrent} e and f, the overlap with the substrate states increases when the substrate states change sign, c.f., Chen's derivative rules for a $p$-wave tip\cite{Chen1990}. The difference in contrast compared to the Cu tip, with a conductance peak over the molecule, can therefore be understood from the overlap of the $p$-wave scattering states from the tip and substrate.

\section{Summary}
The ability to propagate the wave functions into the vacuum gap overcomes the disadvantages of localized atomic orbital DFT in describing the wave functions in the STM gap. This allows us  to use computationally relatively inexpensive DFT calculations to model STM experiments.  The method can furthermore be extended to include \textbf{k}-point sampling, as well as energy dependence to model scanning tunneling spectroscopy. The usefulness of the method is exemplified by the correct description of the CO molecule on Cu(111) depression seen in STM experiments. Here, the depression stems from the sign change of the dominating substrate $s$-wave functions. In contrast, the same molecule shows a protrusion when measured with a CO-functionalized tip, which is due to the $p$-wave character of the tip.

\section{acknowledgement}
We are deeply indebted to Norio Okabayashi and Franz Giessibl for valuable discussions. The computations were performed on resources provided by the Swedish National Infrastructure for Computing (SNIC) at Lunarc. A.G. and M.P. are supported by a grant from the Swedish Research Council (621-2010-3762).
%\bibliographystyle{apsrev}

%\bibliography{/Users/aguadmin/Jobb/PhD/RefPapers/agref.bib}
\end{document}